\documentclass[twocolumn,preprint]{aastex62}
\usepackage{amssymb,amsmath,graphicx,natbib,hyperref}
\usepackage{color}
\usepackage{enumerate}
\usepackage{multirow}

\usepackage[encapsulated]{CJK}
\usepackage[utf8x]{inputenc}
\DeclareGraphicsExtensions{.pdf,.PDF,.png,.PNG,.jpg,.JPG,.jpeg,.JPEG}
\graphicspath{}

\shorttitle{Double nuclei in IC 676}
\shortauthors{Zhou et al.}

\begin{document}
\begin{CJK*}{UTF8}{gbsn}

\title{The Nature of the Double Nuclei in the Barred S0 Galaxy IC676}
\author{Zhimin Zhou(周志民)}
\affiliation{Key Laboratory of Optical Astronomy, National Astronomical Observatories, Chinese Academy of Sciences, Beijing, 100012, China}
\correspondingauthor{Zhimin Zhou}
\email{zmzhou@nao.cas.cn}

\author{Jun Ma}
\affiliation{Key Laboratory of Optical Astronomy, National Astronomical Observatories, Chinese Academy of Sciences, Beijing, 100012, China}
\affiliation{College of Astronomy and Space Sciences, University of Chinese Academy of Sciences, Beijing 100049, China}

\author{Xu Zhou}
\affiliation{Key Laboratory of Optical Astronomy, National Astronomical Observatories, Chinese Academy of Sciences, Beijing, 100012, China}

\author{Hong Wu}
\affiliation{Key Laboratory of Optical Astronomy, National Astronomical Observatories, Chinese Academy of Sciences, Beijing, 100012, China}
\affiliation{College of Astronomy and Space Sciences, University of Chinese Academy of Sciences, Beijing 100049, China}

\begin{abstract}
	The lenticular galaxy IC 676 is a barred galaxy with double nuclei and active star formation in the central region. In this work we present the long-slit spectroscopy and archival multi-wavelength images to investigate the nature and origin of the double nuclei in IC 676. Through photometric 1D brightness profiles and 2D image decomposition, we show that this galaxy consists of a stellar bar with the length of $\sim$ 2.5 kpc and two S\'ersic disks both of which with S\'ersic index $\it n \sim$ 1.3. There is probably little or no bulge component assembled in IC 676. The luminosities of the double nuclei are primarily dominated by young stellar populations within the ages of 1-10 Myr. The northern nucleus has stronger star formation activity than the southern one. The surface densities of the star formation rate in the double nuclei are similar to those in starburst galaxies or the circumnuclear star forming regions in spiral galaxies. Each of the double nuclei in IC 676 likely consists of young massive star clusters, which can be resolved as bright knots in the HST high resolution image. Our results suggest that IC 676 likely has a complex formation and evolutionary history. The secular processes driven by the stellar bar and external accretion may dominate the formation and evolution of its double nuclei. This indicates that the secular evolution involving the internal and external drivers may have an important contribution for the evolution of lenticular galaxies.
\end{abstract}
\keywords{galaxies: individual: IC 676 --- galaxies: elliptical and lenticular, cD --- galaxies: evolution --- galaxies: star formation}

\section{Introduction}
\label{sec:intro}

	Galaxy morphology is fundamental for describing the galaxy properties and inferring their evolution. Of the morphology classifications, lenticular (S0) galaxies provide significant information for studying galaxy evolution. S0s are armless disk galaxies but with smooth appearance like ellipticals and have few signs of star formation and limited amount of gas \citep{Caldwell1993}. They present an intermediate morphology between spirals and elliptical galaxies on the Hubble tuning fork diagram \citep{Hubble1936}. Recently, they are placed in a parallel sequence with the spirals in the revised Hubble diagrams based on their photometric and kinematic properties \citep{van1976,Cappellari2011,Kormendy2012}. These different diagrams depict different evolution scenarios of galaxies. Therefore, studying S0 galaxies is essential to understanding the formation and evolution of the different types of galaxies.

	Despite decades of research, a wide range of open questions remain unanswered with respect to the origin and evolution of S0 galaxies. Recent efforts have shown that S0 galaxies are not a simple transitional class between ellipticals and spirals, instead they unveil a wide range in physical properties such as star formation, stellar populations and kinematics \citep{Gao2018}. Multi-wavelength observations have indicated that 75–80\% of S0 galaxies contain certain amount of dust and gas in general, although much less than spirals \citep{Welch2003}. Evidences of recent star formation have also been found in local S0 galaxies \citep[e.g.,][]{Sil2019}. When compared with elliptical galaxies, their bulge-to-total luminosity ratios B/T can be significantly low \citep{Temi2009, Kormendy2012, Erwin2015}, and a large majority of them contain disc-like pseudobulges in terms of having the S\'ersic indices close to 1--2 \citep{Xiao2016,Mishra2017}.

	S0 galaxies may encompass a wide range of formation mechanisms and evolutionary pathways \citep[e.g.,][]{Fraser2018}. Motivated by the fact that S0 galaxies share similar properties of bulges and disks with spirals, they are thought to be formed from faded spiral galaxies \citep[e.g.,][]{Elmegreen2002,Blanton2009,Laurikainen2013}. Simulations reveal that S0 galaxies in the field can also be transformed from elliptical progenitors by acquiring a disc \citep{Diaz2018}. 

	Observations have shown that S0 galaxies exist in all environments, from dense environments such as clusters and groups to the field, although the fractional population of these galaxies is larger in dense environments than in the field \citep{Wilman2009}. Furthermore, both external and internal processes have important contributions to the evolution of S0 galaxies. Many empirical studies have suggested that S0 galaxies can be formed via a variety of environmental mechanisms including galaxy major/minor mergers \citep{Eliche-Moral2012,Querejeta2015}, ram-pressure stripping \citep{Poggianti2017}, and harassment \citep{Moore1996}. These processes can suppress or quench star formation in galaxies, disrupt the stellar discs, lead to the disappearance of spiral arms, and produce S0-like morphology \citep[e.g.,][]{Rizzo2018,Mishra2019}. 

	Internal secular evolution is also responsible for the transformation of S0s. Stellar bars, one of the important internal drivers, are ubiquitous in discs of S0 and spiral galaxies \citep{Sheth2008,Melvin2014}. Galactic bars are thought to redistribute the angular momentum of gas and stars in galactic disks, induce large-scale streaming motions, and regularize the star formation in galaxies \citep{Kormendy2004,Zhou2015}. In this process, Bars can grow, evolve, and self-destruct \citep{Bournaud2002}. The lenses in almost all S0s are suggested to be formed via bar dissolution, eventually terminating as nearly axisymmetric structures \citep{Kormendy1979}. Another internal driver that could be responsible for the transformation of S0s is feedback processes from active galactic nuclei (AGNs) or supernovae \citep{Penny2018}.

	The nuclear region is an important structural component in galaxies, as it can provide useful information about the formation and evolution of the host galaxies. Especially, the peculiar nuclear morphologies such as double or multiple nuclei deserve closer scrutiny. Galaxies with double nuclei usually reach the stage of galaxy interacting or starburst\citep{Bridge2010,Mezcua2014}, along with certain normal field galaxies \citep[e.g.,][]{Lauer1996,Menezes2018}. Based on spectroscopic investigations, many sources with double nuclei have been identified as AGNs, starburst nuclei, or giant HII regions \citep[e.g.,][]{Mazzarella1993,Moiseev2010,Koss2011}. Galaxy interactions and mergers have been shown to be one of the mechanisms responsible for the generation of double nuclei in disk galaxies \citep{Gimeno2004}. In this scenario, galaxies in the merging phase may manifest double nuclei in their central regions when their outer disks are already mixed with their individual surviving nuclei \citep{Mezcua2014}. Thus single or dual AGNs are tend to be observed in the double nuclei resulting from mergers \citep{Mazzarella2012}. The projected separations of the two nuclei in these systems fall in the range of hundreds to thousands of parsecs based on the merger stages \citep{Mezcua2014}.  
	Beside the mergers, the presence of a nuclear eccentric stellar disk and central supermassive black hole (SMBH) can result in a double nucleus in galaxies \citep{Tremaine1995}. The double nuclei in M31, NGC 4486B and NGC1187 are likely formed through this process along with only parsec-scale separations with their double nuclei \citep{Lauer1996, Lauer2012, Menezes2018}.

	In order to understand the formation and evolution of S0 galaxies, we investigated IC 676, a barred S0 galaxy with double nuclei. This galaxy is at a distance of 20.2 Mpc, corresponding to a scale of $\sim$ 98.2\ pc arcsec$^{\rm -1}$ (NASA Extragalactic Database -- NED). Table \ref{basic_par} gives the global properties of this galaxy.
	IC 676, which appears as early-type morphology, has active star formation in the central region \citep{Contini1998}. This galaxy shows nuclear morphology consistent with the double nuclei \citep{Mazzarella1993}. The projected distance between the double nuclei is about 200 pc along the major axis of the galaxy \citep{Nordgren1995}. Although early studies considered IC 676 as a candidate in the group or pair environment \citep{Peterson1979}, recent observations have suggested that IC 676 is an isolated galaxy \citep{Yuan2010,Cappellari2011}. There is an obvious narrow bar but no evidence for a significant bulge in this system \citep{Erwin2013}. 

	Given the non-typical phenomenon in an isolated early-type galaxy, IC 676 is a promising candidate of S0s because its peculiar properties may provide clues to the agents that drive the evolution and morphology transformation of S0 galaxies. An analysis of the nature of the double nuclei in this galaxy would set important constraints to the physical processes included in galaxy evolution. 

	To analyze and investigate the origin of the double nuclei in IC 676, we coupled our new optical long-slit spectroscopic observations, together with archived images from the Canada France Hawaii Telescope (CFHT) and Hubble Space Telescope (HST). We perform multi-component decomposition to derive its accurate structural parameters, besides analyzing the stellar populations, metallicities, and activity types of the double nuclei. 
	
This paper is organized as follows. Section \ref{sec:data} describes the spectra and archival data that we used to investigate the properties of IC 676. The analysis and results of the structural decompositions and optical spectroscopy are presented in Section \ref{sec:analysis}. We discuss the nature and formation of IC 676's double nuclei in Section \ref{sec:discuss} and provide a brief summary in Section \ref{sec:summary}.

Throughout this paper, we adopt a standard $\Lambda$CDM cosmology with $\it {H}_{\rm 0} {\rm= 70\ km\ s^{-1}\ Mpc^{-1}}$, $\rm {\Omega_M = 0.3}$, and $\rm{\Omega_{\Lambda} = 0.7}$.

\begin{deluxetable}{ccc}
\tablecaption{Global parameters of IC 676
	\label{basic_par}}
	\tablehead{\colhead{Propert} & \colhead{Units} & \colhead{Value}}
\startdata
	R.A.(J2000) & [h:m:s] & 11:12:39.8\\
	Decl.(J2000)& [d:m:s] & +09:03:21.0\\
	Redshift$^a$ &  & 0.00471\\
	Distance$^b$ & [Mpc] & 20.2\\
	Scale$^b$ & [pc/arcsec] & 98.2\\
	D$_{25}$ major axis$^b$ & [arcsec] & 147.3 \\
	$\rm {M_B}^c$ & [mag] & -18.36 \\
	$\rm M_*^d$ & [$\rm M_{\odot}$] & 1.3 $\times$ $\rm 10^9$\\
	SFR$^d$  & [$\rm M_{\odot}\ yr^{-1}$] & 0.28\\ 
	$\rm M_{HI}^e$ & [$\rm M_{\odot}$] & 1.0 $\times$ $\rm 10^8$\\
	$\rm {M_{H_2}}^f$ & [$\rm M_{\odot}$] & 5.1 $\times$ $\rm 10^8$\\
\enddata
	\tablecomments{\\
	$^a$ From SDSS,\\
	$^b$ from NED, \\
	$^c$ from Hyperleda,\\
	$^d$ from \citet{Zhou2018},\\
	$^e$ from \citet{Haynes2011},\\
	$^f$ from \citet{Alatalo2013}.}
\end{deluxetable}

\section{Data Acquisition}
\label{sec:data}

\subsection{Long-slit Spectroscopy}

	Long-slit spectroscopy was performed on 2016 March 15 using the BAO Faint Object Spectrograph and Camera (BFOSC) on the 2.16 m telescope at Xinglong Observatory of the National Astronomical Observatories of China \citep{Fan2016}. The spectra were taken with a slit along the major axis of the stellar bar that covers the two nuclei in the central region simultaneously. The slit width was 2\farcs3 in accordance with the typical local seeing of 2\farcs5. In this observation, the G4, G7 and G8 gratings were used, corresponding to the wavelength range of 3800--8700 \AA, 3780--6760 \AA, and 5800--8280 \AA, giving dispersions of roughly 4.45, 2.13, and 1.79 \AA\ pixel$^{-1}$, respectively. Exposure time for each grating was 1200 $\times$ 3 seconds. A Fe-Ar lamp of each grating was observed for wavelength calibrations. The standard star Feige 34 was observed on the same night for flux calibration. The calibration images, including bias and dome flats, were obtained at the beginning and end of the night. 

The slit spectra were reduced and calibrated using standard CCD procedures in the IRAF environment. The preprocessing included bias subtraction, flat-field correction, and cosmic-ray removal. The spectra were extracted for each grating by employing rectangular apertures on the slit. After extraction, the spectra were wavelength- and flux-calibrated. 


\begin{figure*}
	\centering
	\includegraphics[width=0.7\hsize]{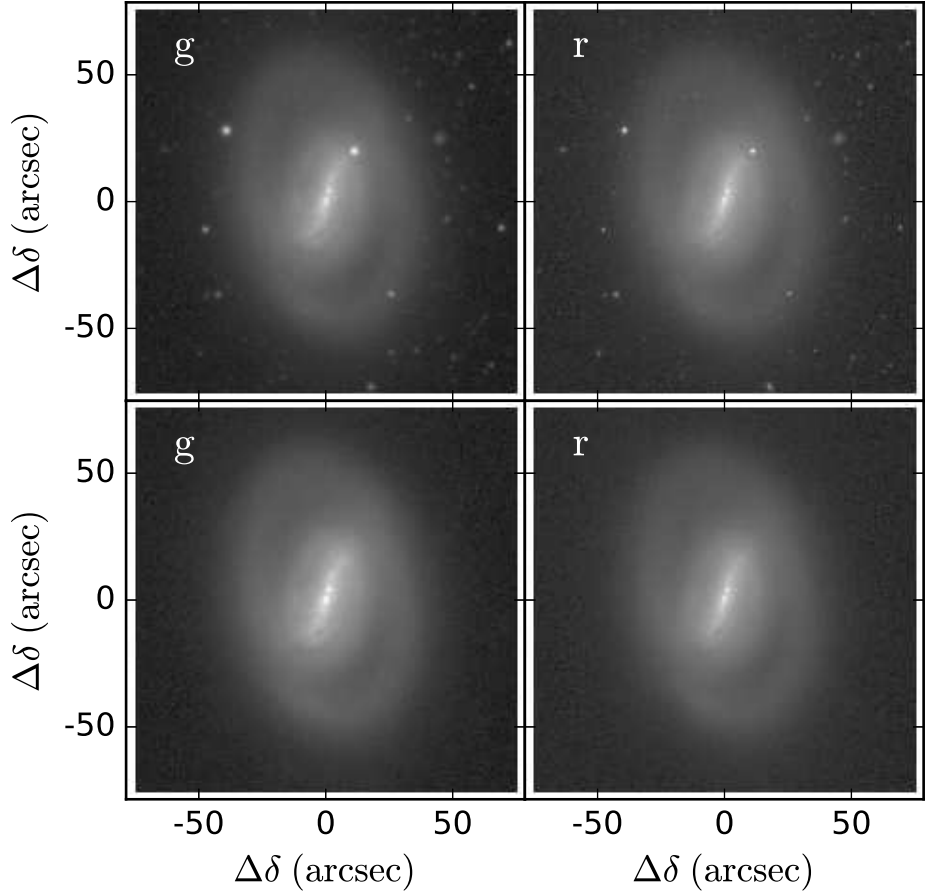}
	\caption{CFHT $\it g$ and $\it r$ images of IC 676. Top: the original images of $\it g$ (left) and $\it r$ (right) band. Bottom: the star-cleaned images of $\it g$ (left) and $\it r$ (right) band.
	\label{fig1}}
\end{figure*}

\begin{figure*}
	\centering
	\includegraphics[width=0.45\hsize]{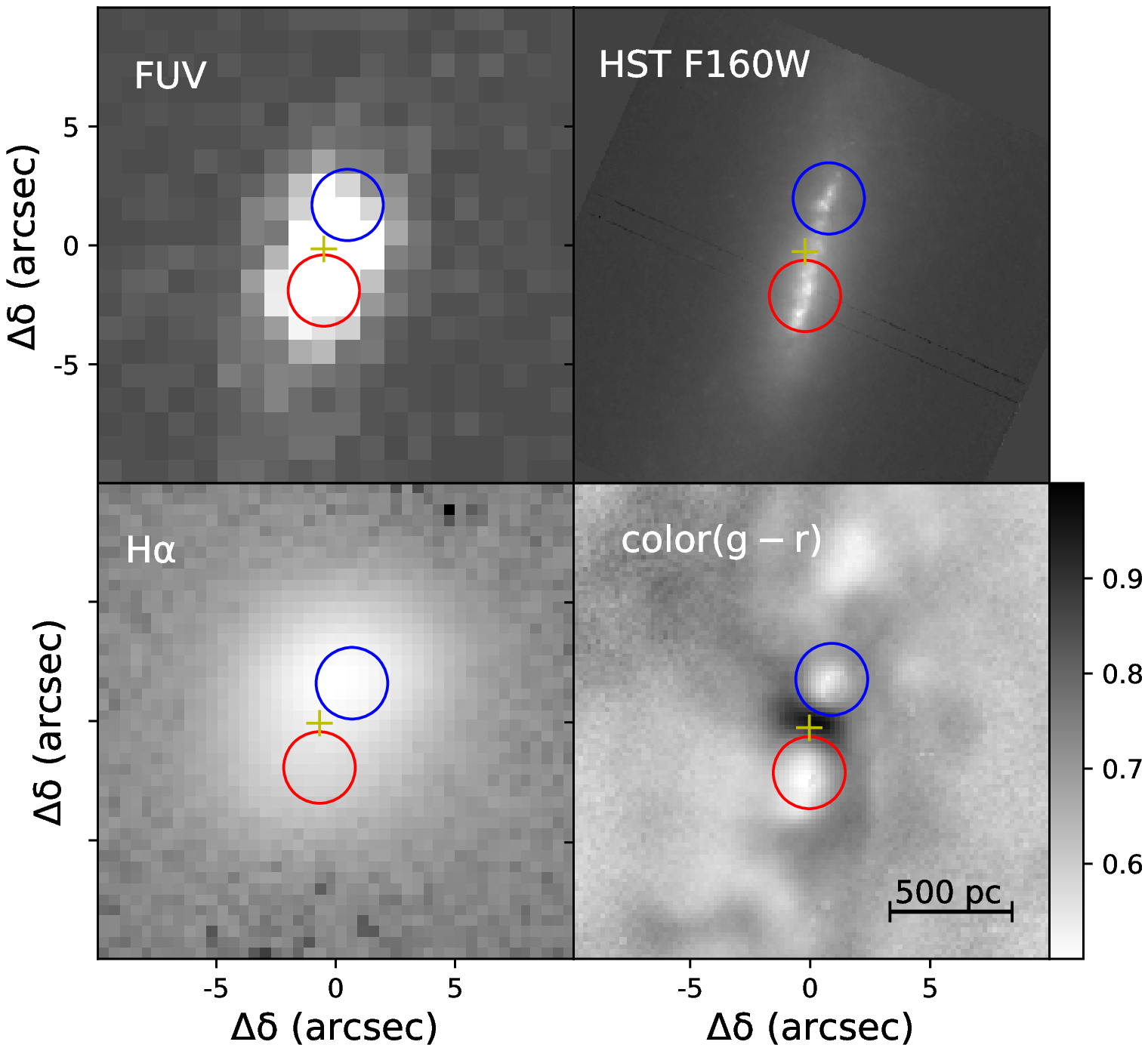}
	\includegraphics[width=0.45\hsize]{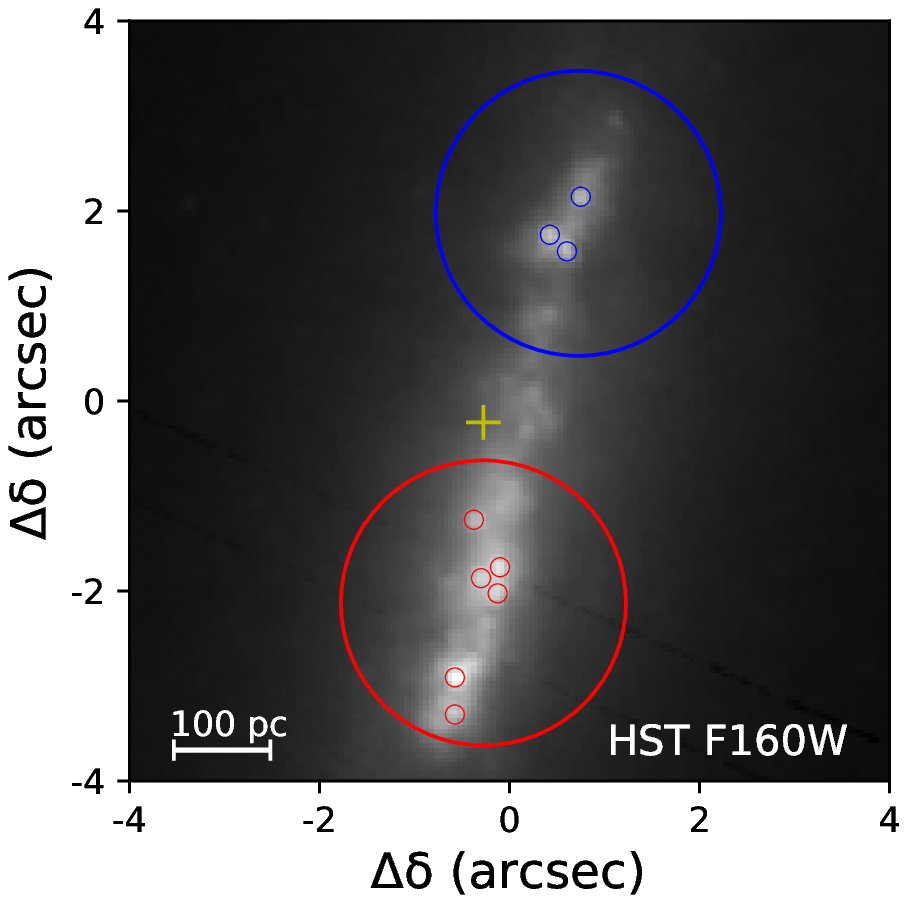}
	\caption{ Central region of IC 676, with North up and East to the left. Left: The images are GALEX FUV, continuum-free H$\alpha$, HST F160W, and CFHT color ($\it g-r$), respectively. The double nuclei are marked with the blue (the northern one) and red (the southern one) circles in each panel. The circles are 1\farcs5 in radius. Right: The zoom-in HST F160W image is shown. The bright knots of the double nucleus regions are marked with small circles (radius $\sim$ 0\farcs1). The yellow cross in all panels marks the photometric centre of the galaxy.
	\label{fig2}}
\end{figure*}

\subsection{Archival data}

Besides our own observations, we also used photometric images from archival sources where the data were of high quality. We searched for archival optical images of IC 676 on the Canadian Astronomy Data Centre (CADC) and selected the images observed with CFHT as part of Proposal \#14BP07 \citep{Duc2015}. These data were obtained with the MegaCam camera on CFHT, and consist of seven g-filter and seven r-filter exposures with the exposure time of 7 $\times$ 345 s for each filter. The raw data were reduced with the pipeline Elixir-LSB and described in \citet{Duc2015}.

We derived the single images after reduction, and combined them for each filter using the AstrOmatic resampling package $\it SWARP$ \citep{Bertin2002, Bertin2010}. The co-added images have a projected pixel size of 0\farcs187 and the full width at half maximum (FWHM) of $\sim$ 1$^{''}$. The surface brightness limits are 28.5-29 mag/arcsec$^2$ in the g-and r-bands. The resulting images are shown in Figure \ref{fig1}.

Besides CFHT images, the far ultraviolet (FUV) image and high-resolution near infrared (NIR) image of IC 676 were retrieved from the Mikulski Archive for Space Telescopes (MAST). This NIR image was observed with the HST NICMOS Camera 2 (NIC2) and F110W filter \citep[HST Proposal \# 11219, PI: Alessandro Capetti;][]{Baldi2010}. This observation was taken on 2008 January 03 with the exposure time of 1151s, and it pointed to the central 19\farcs2 $\times$ 19\farcs2 region of IC 676 with a projected pixel size of 0\farcs075. The FUV image was observed with the $\it Galaxy\ Evolution\ Explorer$ \citep[$\it GALEX$;][]{Martin2005}, and it is characterized by a point-spread function with an FWHM of $\sim$ 6$''$ and a pixel size of 1\farcs5. Furthermore, the continuum-free H$\alpha$ narrow-band image was derived from \citet{Zhou2018}, which was observed with the BFOSC on the 2.16 m telescope at Xinglong Observatory with a resolution of FWHM $=\ 2\farcs5$. Because the FUV and H$\alpha$ emissions are mainly concentrated in the central region of IC 676, Figure \ref{fig2} presents these images for the central 20$''$ $\times$ 20$''$ region of this galaxy in the left panel, and the most inner region from HST image in the right panel.

\subsection{Object Masking and Cleaning}
To obtain accurate results, we removed the signals contributed by unrelated objects, such as foreground bright field stars and background galaxies in CFHT images. The object masking was performed mainly by following the method in \citet{Ho2011} though with certain adjustments. First, we created a mask FITS file that contains the affected pixels. We used $\it SExtractor$ \citep{Bertin1996} to detect the objects in the image and produced the corresponding catalogue and segmentation image. To avoid mistaking the pixels of the main body for the affected ones, we carefully tested the input parameters of $\it SExtractor$ and checked the results visually. 
In addition, we compared the peak values and its background to identify the masking pixels in the region of the target galaxy. In the given parameters of $\it SExtractor$, the initial segmentation image could miss the faint outer halos of the bright affected objects. Thus, we extended the masking region of each object with the growth radius of several to dozens of pixels.  

We used $\it IRAF$ task $\it fitpix$ to clean the contamination from foreground stars and background galaxies based on the object mask created above. This task could replace regions of an image by linear interpolation using the edges across the narrowest dimension. For one image, $\it fitpix$ was performed twice along the lines and columns, respectively. Thereafter, we combined the two fixed images to derive the final star-cleaned image. Figure~\ref{fig1} presents our original and star-cleaned images.

\section{Analysis}
\label{sec:analysis}


\subsection{Morphological analysis}
\label{sec:Structural}
We first visually analyzed the galaxy and its nuclei with the optical maps of the whole galaxy (Figure \ref{fig1}) and the multi-wavelength images of its inner region (Figure \ref{fig2}). IC 676 features an obvious ring, a long stellar bar, and two bright nuclei 2$''$ ($\sim$ 200 pc) apart in projection. The northern nucleus (hereafter Nucleus-N) manifests as a dense core in the optical images, while the southern nucleus (hereafter Nucleus-S) shows more extended and elongated structures. FUV and H$\alpha$ emissions are typical tracers of the star formation activity of galaxies. Hence the FUV and continuum-free H$\alpha$ images indicate that most of the star formation of IC 676 is concentrated in its galactic central region, although these images cannot yet enable to resolve into the double nuclei owing to lower spatial resolution. The colour map $\it g$ - $\it r$ in Figure \ref{fig2} shows that the double nuclei are bluer than other regions of the galaxy, indicating more active star formation.

To further explore the nature of the double nuclei in IC 676, we derived the archival HST F160W image with a high spatial resolution of 0\farcs15. As is evident in Figure~\ref{fig2}, the double nuclei resolve themselves into tens of point sources that are candidate massive star clusters. Nucleus-N primarily consists of three bright clusters in the region of $\sim$ 50 pc. The clusters in Nucleus-S manifest as two groups with a distance of $\sim$ 60 pc, and the northern one is made of 3-5 clusters and the southern one is dominated by one bright cluster. 
We measured the sizes of the clusters to be about 0\farcs12, which is close to the FWHM of the PSF in F160W images. Although we cannot constrain the accurate size of our clusters, we can make sure by considering the effect of PSF that they are securely below $\sim$ 12 pc at a distance of 20.2 Mpc.

\subsection{Structural analysis}
\label{Structural analysis}
\begin{figure*}
	\centering
	\includegraphics[width=0.9\hsize]{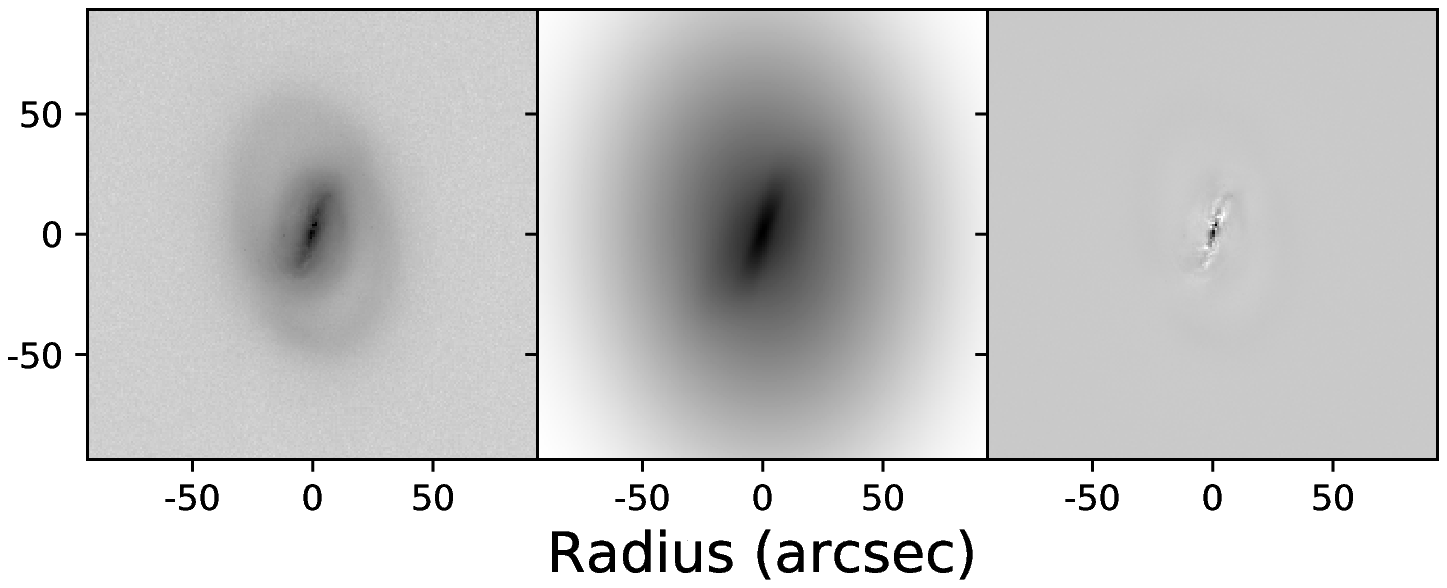}
	\includegraphics[width=\hsize]{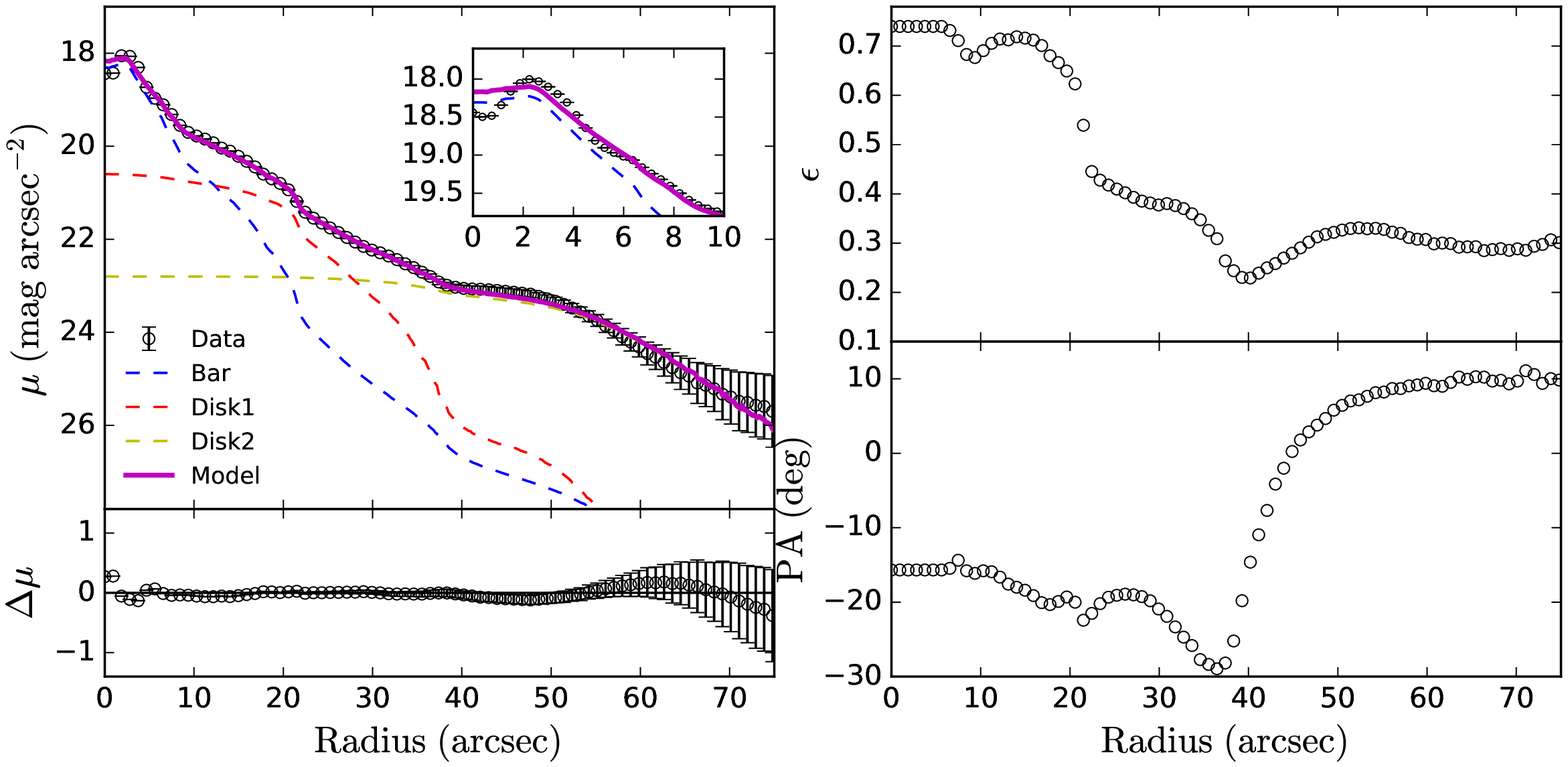}
	\caption{Best-fit 2D models and isophotal analysis of IC 676. The top row shows, from left to right, the star-cleaned $\it r$-band image with gray scale, the 2D model image and the residual image from the 3-components fitting. The left panel on the bottom row shows the radial surface brightness profiles of the date image and the total and individual component models from GALFIT fitting, which are illustrated in the legends. The fitting residual is plotted bellow. The inset panel shows the profile of the galactic central region in 10 arcsec. The right panels on the bottom row show the radial profiles of ellipticity ($\it e$) and position angle (PA). \label{fig3}}
\end{figure*}

We used the $\it IRAF$ task $\it ellipse$ to fit elliptical isophotes of IC 676, and produce the radial surface brightness ($\it \mu$), ellipticity ($\it e$) and position angle (PA) profiles. Because CFHT $\it r$-band image has deeper magnitude limiting and better seeing than the $\it g$-band one, the fitting was only derived on the star-cleaned $\it r$-band image of IC 676. The results are shown in Figure \ref{fig3}.

In the most inner region of the galaxy, an obvious hump can be found in the brightness profile owing to the second nucleus. The bar structure could be identified and measured using these profiles by the method of \citet{Jogee2004}. When a bar exists, the ellipticity steadily increases until it reaches a maximum while the position angle remains constant. Beyond the end of the bar, the ellipticity drop by more than 0.1 and  the position angle changes by more than 5\arcdeg\ in 1-2 arcsecs. Based on the $\it r$-band image of IC 676, a typical stellar bar was detected. The length of the bar is about 21 arcsec ($\sim$ 2.5 kpc) and its ellipticity is 0.70.  

Besides the 1D profiles, we further performed 2D multicomponent decomposition of the $\it r$-band image with $\it GALFIT$ \citep{Peng2002,Peng2010}. We started the fitting with a single S\'ersic disk, which produced a disk component with the S\'ersic index of 2.36 and effective radius of 23\farcs26. With this initial model we obtained the magnitude of the entire galaxy though the stellar bar was missed. Then we added another S\'ersic model for the stellar bar in the subsequent fitting while the outer disk remains in the residual image.
Based on the two-component model, the third S\'ersic disk was added to produce the best fitting, which has given a clean residual for most regions of the galaxy. The structural parameters from the best fittings are presented in Table \ref{table1}. The top panels of Figure \ref{fig3} present the final best fitting results along with the residual image. The 1D surface brightness profiles of the 2D models from our fitting are also plotted in this figure. 

We did not model the two nuclei separately since they, especially the southern nucleus, manifest as compact irregular structures and it is hard to be fitted with a single model. In addition, we masked the double nuclei in our fit to avoid their influence on the fit of the stellar bar. Thus, the components of the double nuclei are presented as the obvious excess of the galactic center in the residual image. Our 2D fitting produces a galactic bar with an effective radius of $\sim$ 8\farcs7 ($\sim$ 1 kpc) and axis ratio of 0.30, which is consistent with the results from 1D profiles. Furthermore, the absence of a bulge component in the fitting models indicates the possibility of either a negligible or nil fraction of the bulge being assembled in IC 676, which is consistent with the results in \citet{Erwin2013}.

Based on the 1D profiles, the photometric centre of IC 676 is located between the double nuclei as marked in Figure~\ref{fig2}. The centres of the three components in our 2D fit are consistent with the 1D result, whereas they have a little offset with each other ($<$0\farcs5).

\begin{deluxetable}{cccccccc}
\tablecaption{Structural Parameters from GALFIT Decomposition \label{table1}}
\tablehead{\colhead{Component} & \colhead{\it m \rm (mag)} & \colhead{$R_e$ ($''$)} & \colhead{\it n} & \colhead{\it b/a} & \colhead{\it PA} & \colhead{$C_0$} & \colhead{$\chi_{\nu}^2$}\\
\colhead{(1)} & \colhead{(2)} & \colhead{(3)} & \colhead{(4)} & \colhead{(5)} & \colhead{(6)} & \colhead{(7)} & \colhead{(8)}}
\startdata
\multicolumn{8}{c}{1-Component Fitting}\\
\hline
	disk  & 12.29 & 23.26 & 2.36 & 0.52 & -17.29 & - & 4.32\\
\hline
\multicolumn{8}{c}{2-Component Fitting}\\
\hline
	bar   & 13.79 & 11.29 & 0.77 & 0.27 & -19.58 & 1.11 & \multirow{2}{*}{2.74} \\
	disk  & 12.69 & 26.42 & 0.94 & 0.75 & -9.64  & - & \\
\hline
\multicolumn{8}{c}{3-Component Fitting}\\
\hline
	bar   & 14.09 & 10.22 & 0.65 & 0.24 & -17.58 & 1.07 &\multirow{3}{*}{2.45}\\
	disk1 & 14.18 & 18.76 & 1.29 & 0.84 & -32.74 & - & \\
	disk2 & 12.85 & 30.69 & 1.00 & 0.71 & 3.26   & - & \\
\enddata
	\tablecomments{Columns: (1): basic components of the GALFIT model. (2): total magnitude of each model for the CFHT r band. (3): effective radius in arcsec. (4): S\'ersic index. (5): axis ratio (b/a) of each component. (6): potion angle measured in degrees East of North. (7): diskyness/boxyness. With $C_0$ $<$ 0, the shape is disky, and conversely, boxy as $C_0$ $>$ 0. (8): reduced $\chi^2$ for the best fit model.}
\end{deluxetable}

\subsection{Spectrum analysis}
\begin{figure*}
	\centering
	\includegraphics[width=0.3\hsize]{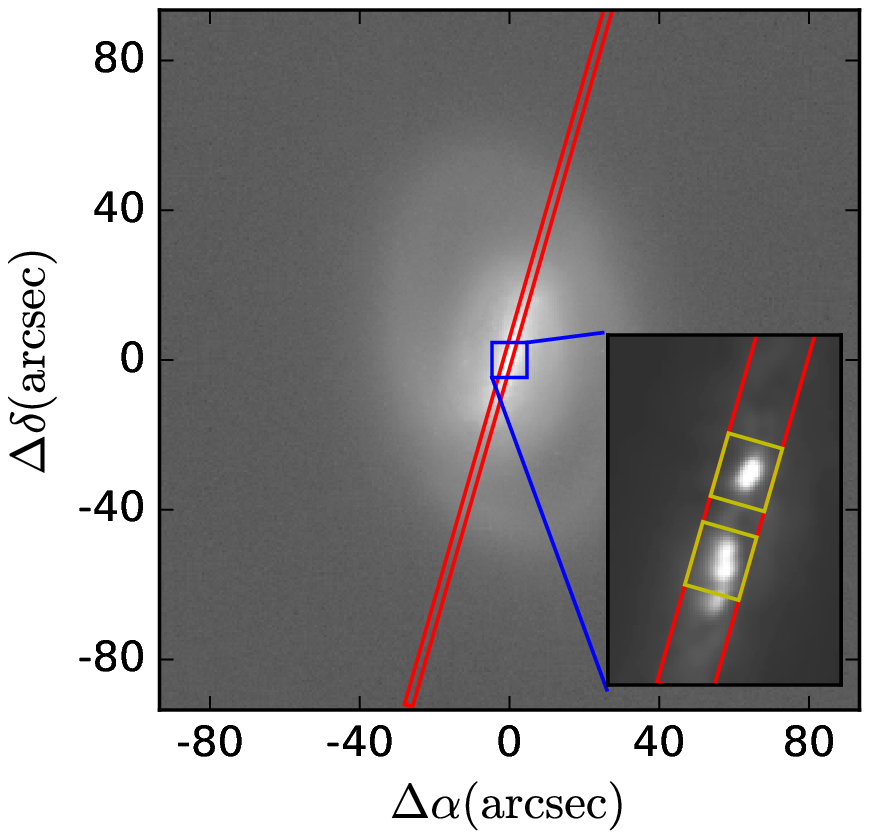}
	\includegraphics[width=0.6\hsize]{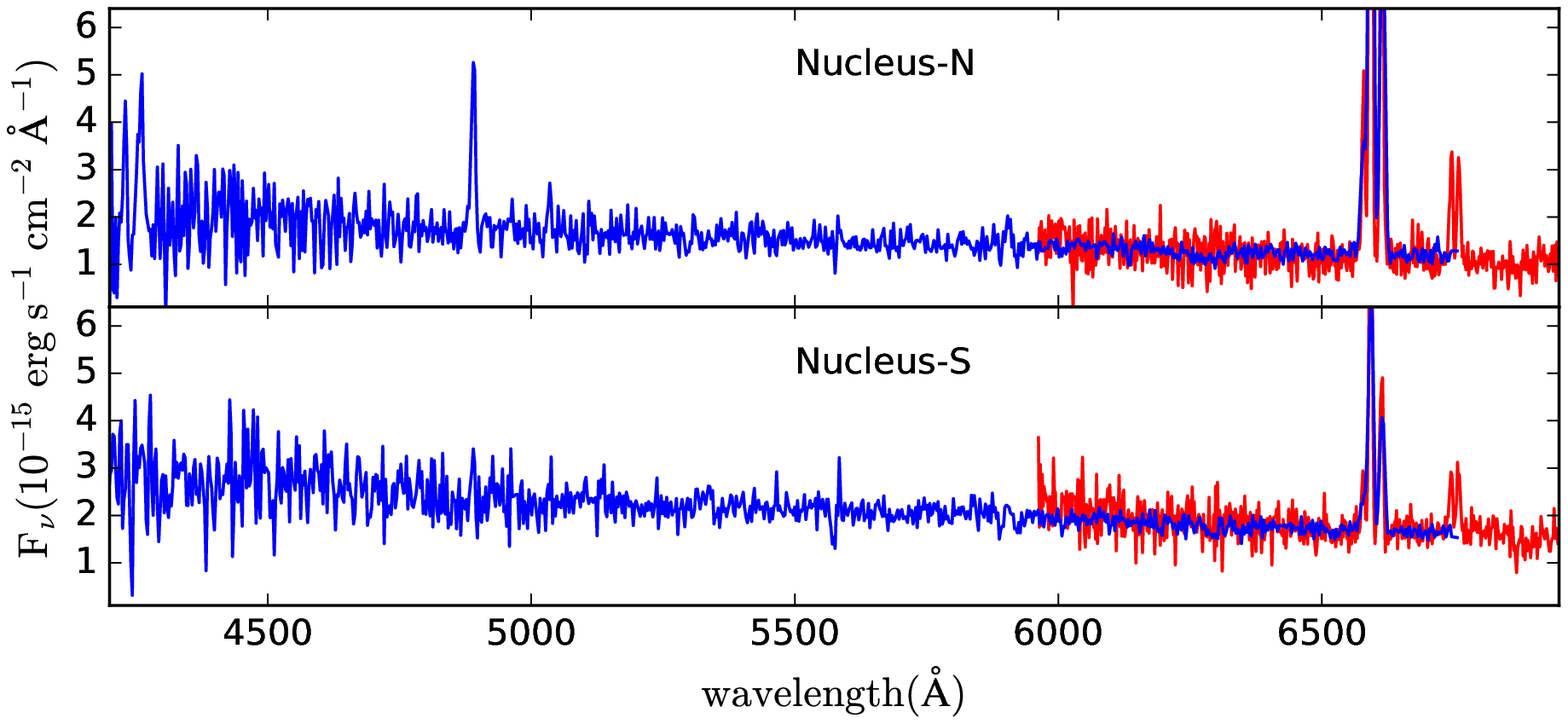}
	\caption{Left: the positioning of the spectroscopic long slit overlapped on $\it r$-band image. The inset shows the nuclei region of IC 676. The rectangles indicated in yellow are the apertures where spectra are extracted in two nuclei. North is up, and east is left. Right: the spectra of the double nuclei. Nucleus-N is the northern one and Nucleus-S is the southern one. The G7- and G8-grating spectra are shown in blue and red, respectively. The G8-grating spectrum is scaled to the flux level of the G7-grating spectrum based on the overlapped wavelength region.
	\label{fig_slit}}
\end{figure*}
\begin{figure*}
	\centering
	\includegraphics[width=0.8\hsize]{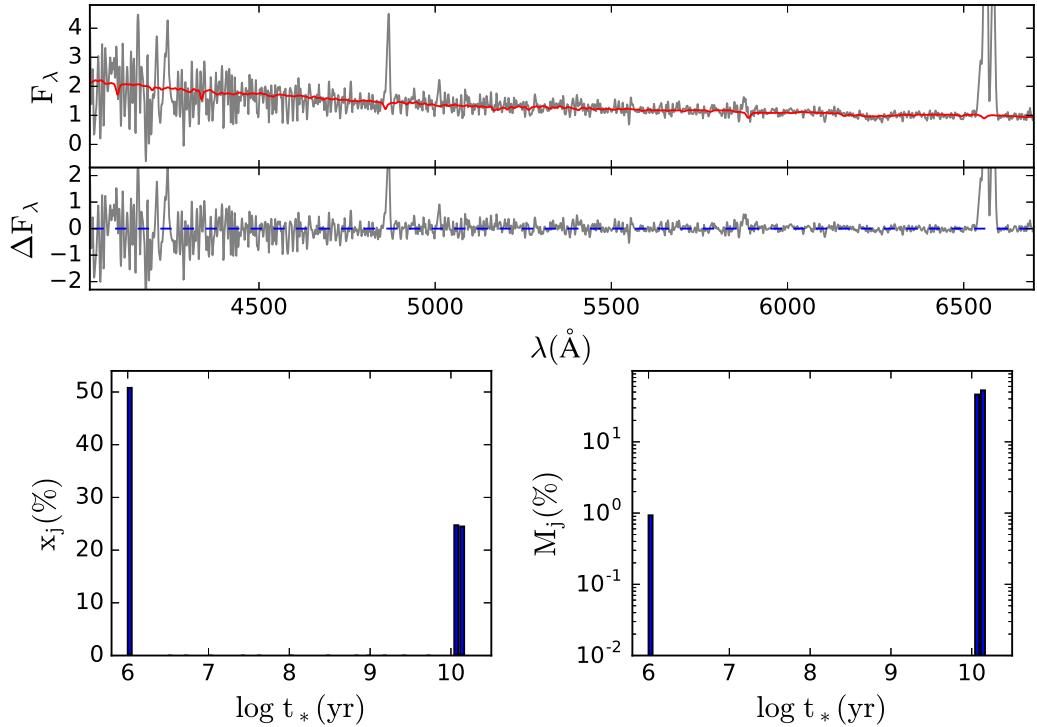}
	\caption{STARLIGHT spectral synthesis for Nucleus-N of IC 676. The top two panels show the observed spectrum (gray), model spectrum (red), and the residual spectrum. The spectra in the two panels are normalized by the flux intensity at 6200.0 \AA. The bottom two panels present the resulting luminosity-weighted (left) and mass-weighted (right) stellar population fractions $\rm x_j$ and $\rm M_j$ as a function of ages of the stellar population templates.
	\label{fig_ssp1}}
\end{figure*}
\begin{figure*}
	\centering
	\includegraphics[width=0.8\hsize]{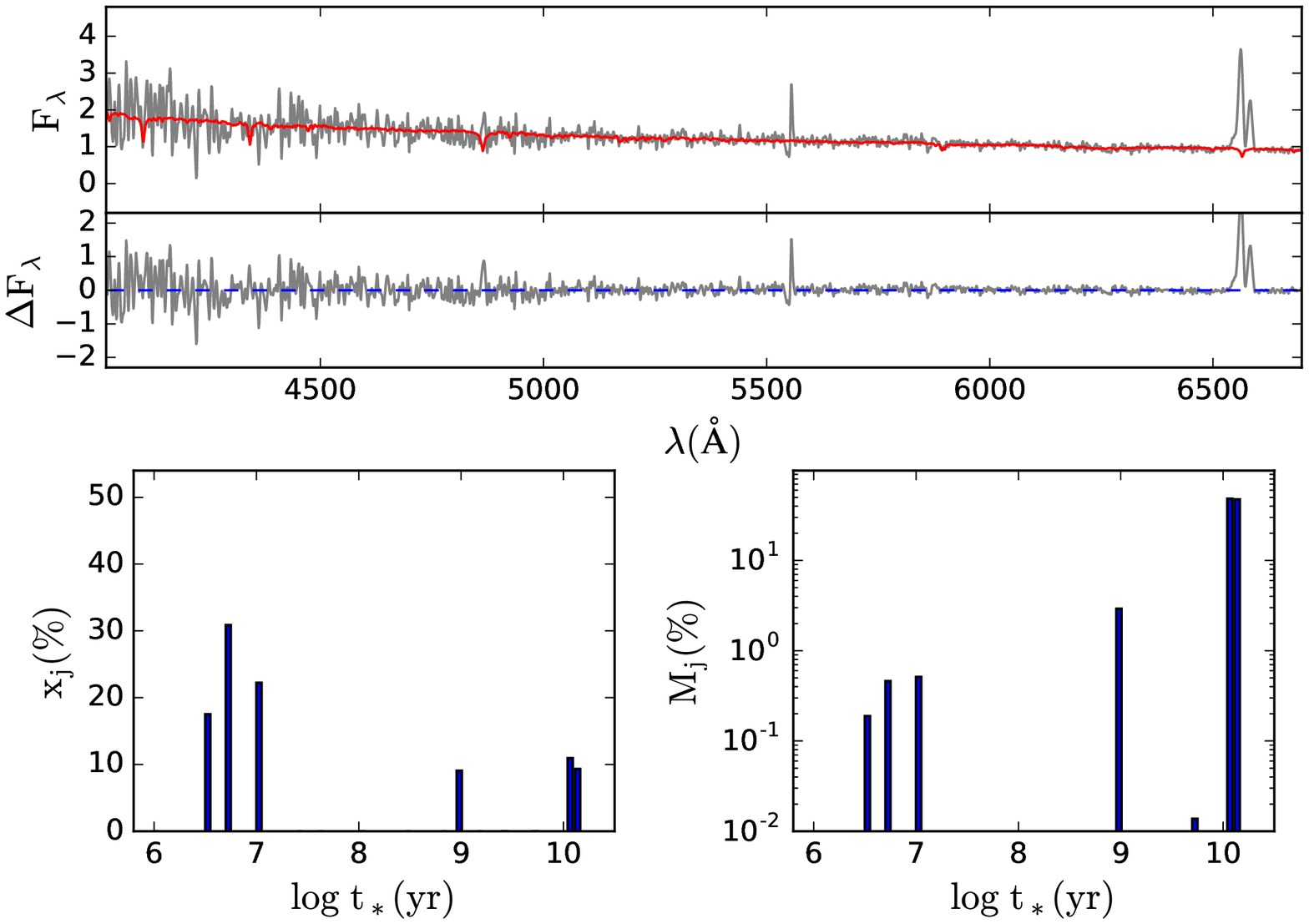}
	\caption{ Same as in Figure \ref{fig_ssp1}, but for spectral synthesis results of Nucleus-S.
	\label{fig_ssp2}}
\end{figure*}

In order to analyze the properties of the two stellar nuclei in IC 676, we extracted the spectra of rectangular apertures on the slit, centered on the two nuclei in the galactic center. The apertures are shown in Figure \ref{fig_slit} and listed in Table \ref{table_ssp}. The size of the apertures is set to 2\farcs3 $\times$ 2\farcs7, corresponding to a physical scale of 283 $\times$ 340 pc$^2$ with the distance of IC 676. In Figure \ref{fig_slit}, we present the extracted spectra. For each object, both the blue- (from G7 grating) and red-side (from G8 grating) spectra are shown, in which the red side is scaled to match the flux level of the blue side.

We applied the spectral synthesis to the extracted spectra by using the STARLIGHT software \citep{Cid2005} and BC03 stellar population model \citep{Bruzual2003}. These template spectra span range in age between 0.001 Gyr to 13.0 Gyr and in metallicity of 0.004 $<$ $\it Z$ $<$ 0.05. The \citet{Cardelli1989} law was adopted for the dust extinction. The blue-side spectra in the wavelength range of 3900-6700\AA\ are used in the fitting, and all the prominent emission lines were masked out during our fitting. We show the results of the spectral synthesis fits in Figures \ref{fig_ssp1} and \ref{fig_ssp2}, together with the fit residuals and the distributions of stellar population components, weighted by luminosity and stellar mass, respectively. The values derived from the fits are summarized in Table \ref{table_ssp}. There are a large fraction of luminosity from the contribution of young stellar populations with ages of 1-10 Myr in both nuclei. Especially for Nucleus-N, about 50$\%$ of the luminosity is dominated by populations with age of about 1 Myr, indicating the much stronger star formation activity recently. Considering the old stellar populations mainly from the background of the galactic central region, the double nuclei are dominated by young stellar populations. We also found that the stellar metallicities of the double nuclei are similar to each other, while Nucleus-N has a higher dust extinction than Nucleus-S.

Based on the spectral synthesis, we also estimated the stellar masses of the star clusters or bright knots in both nuclei (see the HST F160W image in Figure~\ref{fig2}). Assuming 3-5 clusters in each nucleus and removing the old stellar populations from the galactic background, the stellar mass is 1-3 $\times$ $\rm 10^6\ M_{\odot}$ for each cluster. 

\begin{figure*}
	\centering
	\includegraphics[width=0.8\hsize]{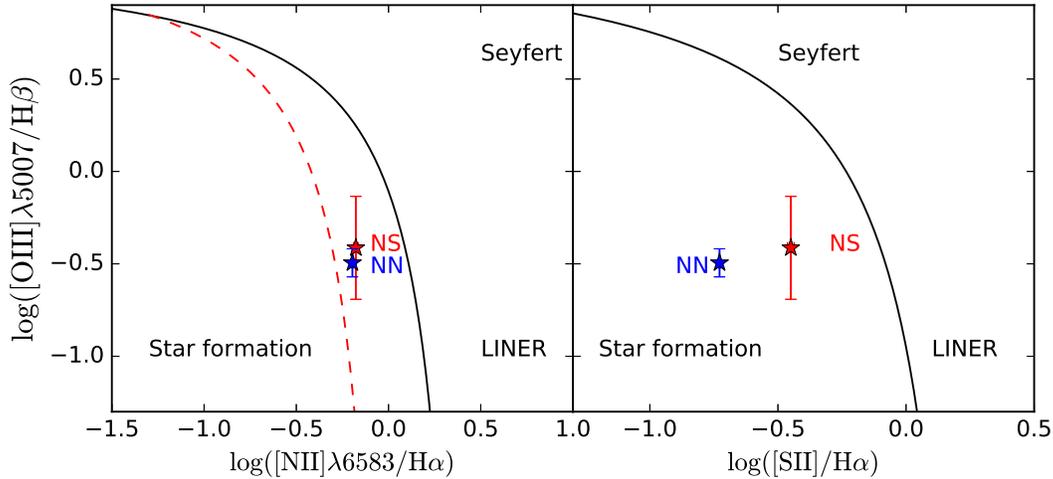}
	\caption{Two BPT optical spectroscopic line ratio diagrams of the double nuclei. Left: the demarcations separating star-forming galaxies and AGNs are from \citet[][; red dashed curve]{Kauffmann2003} and \citet[][; black solid curve]{Kewley2001}. Right: the \citet{Kewley2001} extreme starburst classification line (black solid curve) separates star-forming galaxies from Seyfert and LINER galaxies.
\label{fig_BPT}}
\end{figure*}

To determine the source of gas ionization in the double nuclei, we employ two standard BPT optical spectroscopic diagnostic diagrams \citep{Baldwin1981}, as shown in Figure \ref{fig_BPT}. The fluxes of the optical emission lines in each nucleus were measured, including H$\beta$, [OIII]$\lambda$5007, H$\alpha$, [NII]$\lambda$6584 in the G7-grating spectra and [SII]$\lambda$6717,6731 in the G8-grating spectra. The emission line fluxes of the bar regions were not measured owing to the low signal-to-noise ratio. These fluxes are listed in Table \ref{table_ssp}. Both the nuclei fall into the star formation region in the two BPT diagrams, indicating that the star formation activity has contribution to the photoionization of the nuclei beyond any reasonable doubt. However, both the nuclei are also located in the composite region of star formation and AGN in the left BPT diagram in Figure \ref{fig_BPT}, and thus both nuclei may also harbor AGNs.

The star formation rates (SFRs) of the two nuclei were calculated based on the basis of the H$\alpha$ emission line. The intrinsic extinction correction of H$\alpha$ luminosity was performed with the assumption of \citet{Calzetti2000} extinction law and intrinsic $\rm H\alpha/H\beta$=2.86 for the Case B recombination at a temperature T$_e =$ 10$^4$ K, besides an electron density $\rm n_e = 10^2 cm^{-3}$ \citep{Osterbrock1989}. The SFRs are derived using the calibration from \citet{K98}: $\rm SFR(M_{\odot}yr^{-1})=7.9 \times 10^{-42}[L(H\alpha)](erg~s^{-1})$, where $\rm L(H\alpha)]$ is the extinction-corrected H$\alpha$ luminosity. With the assumption that all $\rm H\alpha$ emission in IC 676 is from the star formation activity, we obtained SFRs of 0.19 $\rm M_{\odot}yr^{-1}$ and 0.08 $\rm M_{\odot}yr^{-1}$, corresponding to 3.21 and 1.40 $\rm M_{\odot}yr^{-1}kpc^{-2}$ for Nucleus-N and Nucleus-S, respectively. This confirms more active star formation in Nucleus-N than in Nucleus-S. In addition, these values are similar to SFR surface densities of starburst galaxies or the circumnuclear star-forming regions in spiral galaxies \citep{K98, Kennicutt2012}. The total SFR of IC 676 is 0.28 $\rm M_{\odot}yr^{-1}$ based on the measurement of \citet{Zhou2018}. Therefore, nearly all of the star formation activity in this galaxy is concentrated in the double nuclei, consistent with our finding displayed in Figure \ref{fig2}.
\begin{deluxetable}{lcc}
\tablecaption{Emission-line and Stellar Population Parameters for two nuclei of IC 676 
	\label{table_ssp}}
	\tablehead{\colhead{Parameter} & \colhead{Nucleus-N} & \colhead{Nucleus-S}}
\startdata
	\multicolumn{3}{l}{Position} \\
	R.A.(J2000) & 11:12:39.7 & 11:12:39.8\\ 
	Decl.(J2000)& +09:03:24.83 & +09:03:21.30\\
\hline
	\multicolumn{3}{l}{Stellar Population$^a$ (light-weighted)}\\
	Young ($\%$) & 50.8 & 70.6\\
	Median ($\%$) & 0.0 &9.1\\
	Old ($\%$) & 49.2 &20.3\\
\hline
	\multicolumn{3}{l}{Emission-line flux ($\rm 10^{-15}~erg~s^{-1}~cm^{-2}$)}\\
	H$\beta$                & 7.7$\pm$0.3  & 2.5$\pm$0.5\\
	H$\alpha$               & 62.0$\pm$0.2 & 21.9$\pm$0.3\\
    $\rm [O~III]\lambda$5007     & 2.5$\pm$0.4  & 1.0$\pm$0.6\\
	$\rm [N~II]\lambda$6584      & 39.5$\pm$0.2 & 14.5$\pm$0.4\\
	$\rm [S~II]\lambda$6717,6731 & 11.6$\pm$0.2 & 7.8$\pm$0.4\\
\enddata
\tablecomments{a: the percentage of stellar population with different ages weighted by luminosity. Young is the population with age $\rm <$100 Myr, 100 Myr - 1 Gyr for Median, and age $\rm >$1 Gyr for Old.}
\end{deluxetable}

\section{Discussion}
\label{sec:discuss}

\subsection{The nature of the double nuclei in IC 676}

In the previous section, we have inferred the physical properties of the double nuclei in S0 barred galaxy IC 676. The residual image of the GALFIT 2D decomposition revealed certain peculiar morphological features of these nuclei. We found that Nucleus-S manifested more extended and elongated structures while that of Nucleus-N presents as a dense core in our optical images. The spectrum analysis indicated that there are a large percentage of young stellar populations in the two nuclei and both nuclei have active star formation.  

Furthermore, the double nuclei can be resolved into several massive star clusters with parsec-scale size in the HST image. We assume that these clusters in IC 676 are young massive star clusters (YMCs). YMCs are abundant in interacting gas-rich galaxies and dwarf starbursts \citep{Cao2007, Portegies2010}. YMCs are typically found to be within 20 pc with the mass of $\rm 10^5-10^6\ M_{\odot}$ \citep{Larsen1999,Portegies2010}. YMCs appear to be formed in giant molecular clouds (GMCs), and they are often associated with violent star formation activity \citep{Portegies2010}. Thus, a large number of YMCs are formed younger than about 100 Myr, which is consistent with the starburst age of their host galaxies \citep{de2013}. 

The photometry from CFHT and HST images of IC 676 has demonstrated that the size of each bright knot in the double nuclei is below 12 pc. The spectrum analysis revealed the young stellar population with ages of $\sim$ $10^6$ - $10^8$ yr in the double nuclei (Figures~\ref{fig_ssp1} and \ref{fig_ssp2}). The FUV and H$\alpha$ images of IC 676 present typical giant HII regions in the galactic center, while these images can not yet allow us to resolve the double nuclei owing to less well spatial resolution (Figure \ref{fig2}). Furthermore, the CO-integrated intensity map observed by the IRAM 30-m telescope shows intense molecular gas accumulated in the nuclei region of IC 676 \citep{Alatalo2013}. Therefore, most of the properties of the knots in the double nuclei are consistent with the characters of the YMCs that we have mentioned above.

We also considered the probability that the bright knots in the double nuclei are nuclear star clusters (NSCs). NSCs are dense stellar systems in galaxy centres, and they are present in a wide variety of galaxies across the entire Hubble sequence \citep{den2014, Carson2015}. Based on the criteria of \citet{Boker2002}, NSC is often the brightest and only cluster within a kiloparsec from the photometric center of the host galaxy. However, in the sources from this study, there is not a single dominant, isolated point-like source that can be detected at or close to the galaxy center. Therefore, we did not classify our clusters as NSCs.

\subsection{Possible formation scenarios of IC 676's nuclei}

We have shown above that there is active star formation in the galactic center of IC 676, and its double nuclei consist of several bright knots, which are identified as YMCs. Moreover, our recent work have derived that IC 676 has a stellar mass of 1.3 $\times$ $\rm 10^9 ~M_{\odot}$, an extinction-corrected H$\alpha$ star formation rate (SFR) of 0.28 $\rm M_{\odot}~yr^{-1}$, and neutral hydrogen (HI) gas mass of 1.0 $\times$ $\rm 10^8~ M_{\odot}$ \citep{Zhou2018}. Thus, the star formation efficiency (SFE) of neutral gas is $\rm 10^{-8.56}~yr^{-1}$ (i.e., 2.7 $\times$ $\rm 10^{-9}~yr^{-1}$), obviously higher than the mean SFE of $\rm 10^{-9.95}~yr^{-1}$ derived for the ALFALFA HI-selected galaxies by \citet{Huang2012}. The total mass of molecular gas $\rm H_2$ measured from CO observation is 5.13 $\times$ $\rm 10^8~ M_{\odot}$, taken from \citet{Alatalo2013}. Thus, the SFE of total gas is $\rm \sim~10^{-9.34}~yr^{-1}$ for IC 676, similar to that of normal disk galaxies \citep{K98}.

As mentioned before, IC 676 is a barred S0 galaxy. However, such phenomena as above appear to be non-typical in this type of galaxy. The factors that trigger the star formation in this galaxy, and the reason behind presence of two nuclei in the galactic center need to be understood. Major merger is an important driver for the evolution of S0 galaxies \citep{Prieto2013} and YMCs in galaxies \citep{Kruijssen2012}. However, major merger could be excluded in the recent physical processes of IC 676. 
Although IC 676 was previously classified as the merger candidate \citep{Gimeno2004}, optical images of IC 676 show no indication of tidal features resulting from galaxy interactions (Figure~\ref{fig1}). While it is also plausible that this galaxy might be in the post-merger stage of a merger event when the features of galaxy interaction have already vanished \citep{Mezcua2014}. IC 676 is located in the outside of Virgo Cluster with a projected distance of about 7.5 Mpc to the cluster centre, much larger than the radius of the cluster. Based on the MPA-JHU spectroscopic sample of Sloan Digital Sky Survey \citep[SDSS;][]{Brinchmann2004,York2000}, the nearest galaxy of IC 676 is CGCG 039-068 with a distance between each other of $\sim$ 0.5 Mpc which is significantly larger than the virial radii of both galaxies. Thus, the effects on IC 676 from other galaxies can be ignored.

Furthermore, there is a large-scale stellar bar and two regular rings (inner and outer) in the disk of IC 676. Given that the bar is long-lived, this galaxy must be free of major or even minor mergers for a long time, since merger events could scramble disks and destroy the bar and rings \citep{Kormendy2004}.
Moreover, the inner and outer rings in IC 676 are the features related to specific orbital resonances with the pattern speed of the stellar bar \citep{Buta2017}. They are considered to be the result of secular evolution processes from long-lived non-axisymmetries that redistribute the material and angular momentum in discs \citep{Comeron2014, Buta2015}. Besides that, there is not a bulge component found in IC 676, especially the classical bulge which is assumed to be the product of the major merger \citep{Athanassoula2005}. These features in IC 676 completely ruled out a recent merger event.

Another possible driver of the galaxy evolution in IC 676 is the stellar bar. In Section~\ref{sec:Structural}, our images and structural analysis have indicated the existence of a strong bar in IC 676 (Figure~\ref{fig3}). Multi-wavelength observations have also shown that strong bars are common in discs of S0 galaxies \citep{Laurikainen2007,Lansbury2014}. Lots of Numerical simulations and multi-wavelength observations have established that the gas in disks can lose its angular momentum in the effect of stellar bars and then move inward towards the inner region of galaxies followed by a trigger for active star formation in the dense molecular gas reservoirs \citep[e.g.,][]{Athanassoula2013,Consolandi2017}. Radio CO observations have indicated that IC 676 has high molecular gas concentrations in its central kiloparsecs, and the gas is distributed along the bar region \citep[see Figure A1. in ][]{Alatalo2013}. This may indicate the bar-driven gas inflow in IC 676. Furthermore, the star formation activity and the young stellar populations in the central region of IC 676 is also consistent with the outcome of bar-driven secular evolution.

Given that the material is redistributed by the stellar bar in the galactic disk, the external origin of the material fueling the star formation is still not clear. Various studies have suggested that the external mechanisms such as cosmological accretion and minor mergers can bring in gas and resume the formation of stars in early-type galaxies \citep[e.g.,][]{Davis2015,Bryant2019}. A key signature of the external gas replenishment is that the angular momentum of the gas is misaligned from the stars in the galaxy \citep{Davis2019}. For IC 676, \citet{Davis2011} measured its kinematic position angles (PAs) of gas and stars, and found that the ionized gas are likely to be kinematically misaligned for the stars with $\rm \Delta PA = 30.5\arcdeg \pm 29.6\arcdeg$, while its molecular gas seems to be kinematically aligned with respect to the stars ($\rm \Delta PA = 2\arcdeg \pm 15.8\arcdeg$). However, based on the large uncertainty of the position angles, we cannot ascertain the effect of the external  accretion of the gas. Furthermore, the ionized gas kinematic PA of IC 676 is probably affected by its stellar bar \citep{Davis2011}.

Additionally, the offset of the ionized gas component in IC 676 is also probably heated and influenced by AGN activity. Many studies have confirmed that AGN feedback can drive gas outflows and affect star formation in host galaxies \citep[e.g.,][]{Ishibashi2013,Manzano2019}. While there is not a clear sign of an AGN in IC 676 on the BPT diagram (Figure~\ref{fig_BPT}), both nuclei are located on the composite region of AGN and star formation. Thus, we can not rule out IC 676 as an AGN host, and AGN activity may also make a contribution to the star formation and evolution of this galaxy. 

As shown in Figure~\ref{fig2}, the double nuclei in IC 676 are two groups of YMCs in the galactic central region, which are different from the double or multi-nuclei in the merging systems. This indicates that the evolution of IC 676's nuclei are governed by secular processes rather than by major mergers. Furthermore, there is not a (classical) bulge component found in this galaxy, and the double nuclei are likely a part of the pseudobulge of IC 676, or will transform to the pseudobulge in the future. The bar-driven gas inflow makes the dense molecular gas reservoir in the galactic central region, and then it triggers the formation of YMCs. The radio observation shows that the 5 GHz continuum emission in IC 676 is mainly concentrated on the northern nucleus, and characterized by non-axisymmetric, extended morphology, which could be the result of SF, or a mixture of AGN and SF \citep{Nyland2016}. It is interesting that negligible emission is coming from the southern nucleus. Maybe that the gas on the southern nucleus are (mostly) consumed by SF, or be dispelled by some physical processes such as feedbacks from AGN or supernova.
Besides that, the presence of a nuclear eccentric stellar disk can also result in the double galactic nuclei such as M31 \citep{Lauer2012}. Although we could not confirm the existence of an eccentric stellar and gas rotating nuclear disk in IC 676 based on our current data, this opens the possibility for the formation of the double nuclei.

Overall, our results suggest that the barred S0 galaxy IC 676 likely has a complex formation and evolutionary history. Various mechanisms, especially the secular processes driven by the stellar bar and external accretion, provide contributions to the formation of the double nuclei and evolution of IC 676. This indicates that the secular evolution involving the internal and external mechanisms may have significant contribution to the formation and evolution of lenticular galaxies.


\section{Summary}
\label{sec:summary}

In this paper we utilize available multi-wavelength images and optical long-slit spectra to investigate the nature of the double nuclei in the isolated barred lenticular galaxy, IC 676. We perform detailed 1D profiles and 2D decompositions of this galaxy’s CFHT $\it r$-band data of this galaxy, and apply the spectral synthesis to analyze the stellar populations and activity types of the double nuclei. Our primary results can be summarized as follows.

\begin{enumerate}
	\item{} Based on 1D brightness profiles and 2D image decompositions, the best-fit 2D model for IC 676 consists of one bar component and two S\'ersic disks with S\'ersic index $\it n\ \sim$ 1.3. The stellar bar in IC 676 has the length of $\sim$ 2.5 kpc and axis ratio of 0.30. There is probably little or no bulge component assembled in IC 676.

	\item{} The luminosities of the double nuclei are mainly dominated by young stellar populations with ages of 1-10 Myr. Compared with the southern nucleus, the northern one has stronger star formation activity and higher dust extinction, though with similar stellar metallicity. BPT diagrams indicate that the star formation activity is the main source of gas ionization in the double nuclei, while certain contribution may also come from AGNs.

	\item{} Nearly all of the star formation in IC 676 is concentrated on the double nuclei. The northern nucleus has more than twice the SFR surface density of the southern one. Both of them are similar to those in the starburst galaxies or the circumnuclear star-forming regions in spiral galaxies.

\item{} We have discussed the nature of the double nuclei, and suppose that each of them consists of several YMCs, which can be resolved as bright knots in the HST high resolution image. The secular processes may primarily contribute to the formation and evolution of the double nuclei. 
	The gas in IC 676 is likely affected by the stellar bar, and driven into the inner regions of the galaxy, resulting in the star formation activity. The interference of galaxy interaction is likely to be negligible.
This indicates that the secular evolution involving the internal or external drivers may have an important contribution for the evolution of lenticular galaxies.

\end{enumerate}

\acknowledgements
\label{sec:acknow}
We thank the anonymous referee for numerous critical comments and instructive suggestions, which have significantly improved both the content and presentation of this paper.
This work was supported by the Chinese National Natural Science Foundation grands No. 11890693, 11433005, 11873053, and 11673027, 11733006, by the External Cooperation Program of the Chinese Academy of Sciences (grant no. 114A11KYSB20160057), and by National Key R\&D Program of China No. 2019YFA0405501.

This research used the facilities of the Canadian Astronomy Data Centre operated by the National Research Council of Canada with the support of the Canadian Space Agency.
This work is based on observations made with the $\it Galaxy\ Evolution\ Explorer$ ($\it GALEX$). $\it GALEX$ is operated for NASA by the California Institute of Technology under NASA contract NAS5-98034. 
This research is based on observations made with the NASA/ESA $\it Hubble\ Space\ Telescope$ obtained from the Space Telescope Science Institute, which is operated by the Association of Universities for Research in Astronomy, Inc., under NASA contract NAS 5–26555. These observations are associated with $\it HST$ Proposal \# 11219.
This research has also made use of the NASA/IPAC Extragalactic Database (NED) which is operated by the Jet Propulsion Laboratory, California Institute of Technology, under contract with the National Aeronautics and Space Administration.

\bibliographystyle{aasjournal}
\bibliography{}

\begin{thebibliography}{}

\bibitem[Alatalo et al.(2013)]{Alatalo2013} Alatalo, K., Davis, T.~A., Bureau, M., et al.\ 2013, \mnras, 432, 1796
\bibitem[Athanassoula(2005)]{Athanassoula2005} Athanassoula, E.\ 2005, \mnras, 358, 1477
\bibitem[Athanassoula et al.(2013)]{Athanassoula2013} Athanassoula, E., Machado, R.~E.~G., \& Rodionov, S.~A.\ 2013, \mnras, 429, 1949
\bibitem[Baldi et al.(2010)]{Baldi2010} Baldi, R.~D., Chiaberge, M., Capetti, A., et al.\ 2010, \apj, 725, 2426
\bibitem[Baldwin et al.(1981)]{Baldwin1981} Baldwin, J. A., Phillips, M. M., \& Terlevich, R. 1981, \pasp, 93, 5
\bibitem[Bertin(2010)]{Bertin2010} Bertin, E.\ 2010, SWarp: Resampling and Co-adding FITS Images Together, ascl:1010.068
\bibitem[Bertin et al.(2002)]{Bertin2002} Bertin, E., Mellier, Y., Radovich, M., et al.\ 2002, Astronomical Data Analysis Software and Systems XI, 228
\bibitem[Bertin, \& Arnouts(1996)]{Bertin1996} Bertin, E., \& Arnouts, S.\ 1996, \aaps, 117, 393
\bibitem[Blanton, \& Moustakas(2009)]{Blanton2009} Blanton, M.~R., \& Moustakas, J.\ 2009, \araa, 47, 159
\bibitem[B{\"o}ker et al.(2002)]{Boker2002} B{\"o}ker, T., Laine, S., van der Marel, R.~P., et al.\ 2002, \aj, 123, 1389 
\bibitem[Bournaud, \& Combes(2002)]{Bournaud2002} Bournaud, F., \& Combes, F.\ 2002, \aap, 392, 83
\bibitem[Brinchmann et al.(2004)]{Brinchmann2004} Brinchmann, J., Charlot, S., White, S. D. M., et al. 2004, \mnras, 351, 1151      
\bibitem[Bridge et al.(2010)]{Bridge2010} Bridge, C.~R., Carlberg, R.~G., \& Sullivan, M.\ 2010, \apj, 709, 1067
\bibitem[Bruzual, \& Charlot(2003)]{Bruzual2003} Bruzual, G., \& Charlot, S.\ 2003, \mnras, 344, 1000
\bibitem[Bryant et al.(2019)]{Bryant2019} Bryant, J.~J., Croom, S.~M., van de Sande, J., et al.\ 2019, \mnras, 483, 458
\bibitem[Buta(2017)]{Buta2017} Buta, R.~J.\ 2017, \mnras, 470, 3819
\bibitem[Buta et al.(2015)]{Buta2015} Buta, R.~J., Sheth, K., Athanassoula, E., et al.\ 2015, \apjs, 217, 32
\bibitem[Caldwell et al.(1993)]{Caldwell1993} Caldwell, N., Rose, J.~A., Sharples, R.~M., et al.\ 1993, \aj, 106, 473
\bibitem[Calzetti et al.(2000)]{Calzetti2000} Calzetti, D., Armus, L., Bohlin, R.~C., et al.\ 2000, \apj, 533, 682
\bibitem[Cao, \& Wu(2007)]{Cao2007} Cao, C., \& Wu, H.\ 2007, The Astronomical Journal, 133, 1710
\bibitem[Cappellari et al.(2011)]{Cappellari2011} Cappellari, M., Emsellem, E., Krajnovi{\'c}, D., et al.\ 2011, \mnras, 416, 1680
\bibitem[Cardelli et al.(1989)]{Cardelli1989} Cardelli, J.~A., Clayton, G.~C., \& Mathis, J.~S.\ 1989, \apj, 345, 245
\bibitem[Carson et al.(2015)]{Carson2015} Carson, D.~J., Barth, A.~J., Seth, A.~C., et al.\ 2015, \aj, 149, 170 
\bibitem[Cid Fernandes et al.(2005)]{Cid2005} Cid Fernandes, R., Mateus, A., Sodr{\'e}, L., et al.\ 2005, \mnras, 358, 363
\bibitem[Comer{\'o}n et al.(2014)]{Comeron2014} Comer{\'o}n, S., Salo, H., Laurikainen, E., et al.\ 2014, \aap, 562, A121
\bibitem[Consolandi et al.(2017)]{Consolandi2017} Consolandi, G., Dotti, M., Boselli, A., et al.\ 2017, \aap, 598, A114
\bibitem[Contini et al.(1998)]{Contini1998} Contini, T., Considere, S., \& Davoust, E.\ 1998, \aaps, 130, 285
\bibitem[Davis et al.(2011)]{Davis2011} Davis, T.~A., Alatalo, K., Sarzi, M., et al.\ 2011, \mnras, 417, 882
\bibitem[Davis et al.(2015)]{Davis2015} Davis, T.~A., Rowlands, K., Allison, J.~R., et al.\ 2015, \mnras, 449, 3503
\bibitem[Davis, \& Young(2019)]{Davis2019} Davis, T.~A., \& Young, L.~M.\ 2019, arXiv e-prints, arXiv:1909.01230
\bibitem[de Grijs et al.(2013)]{de2013} de Grijs, R., Anders, P., Zackrisson, E., et al.\ 2013, Monthly Notices of the Royal Astronomical Society, 431, 2917
\bibitem[den Brok et al.(2014)]{den2014} den Brok, M., Peletier, R.~F., Seth, A., et al.\ 2014, \mnras, 445, 2385 
\bibitem[Diaz et al.(2018)]{Diaz2018} Diaz, J., Bekki, K., Forbes, D.~A., et al.\ 2018, \mnras, 477, 2030
\bibitem[Duc et al.(2015)]{Duc2015} Duc, P.-A., Cuillandre, J.-C., Karabal, E., et al.\ 2015, \mnras, 446, 120
\bibitem[Eliche-Moral et al.(2012)]{Eliche-Moral2012} Eliche-Moral, M.~C., Gonz{\'a}lez-Garc{\'\i}a, A.~C., Aguerri, J.~A.~L., et al.\ 2012, \aap, 547, A48
\bibitem[Elmegreen et al.(2002)]{Elmegreen2002} Elmegreen, D.~M., Elmegreen, B.~G., Frogel, J.~A., et al.\ 2002, \aj, 124, 777
\bibitem[Erwin, \& Debattista(2013)]{Erwin2013} Erwin, P., \& Debattista, V.~P.\ 2013, \mnras, 431, 3060
\bibitem[Erwin et al.(2015)]{Erwin2015} Erwin, P., Saglia, R.~P., Fabricius, M., et al.\ 2015, \mnras, 446, 4039
\bibitem[Fan et al.(2016)]{Fan2016} Fan, Z., Wang, H., Jiang, X., et al.\ 2016, \pasp, 128, 115005 
\bibitem[Fraser-McKelvie et al.(2018)]{Fraser2018} Fraser-McKelvie, A., Arag{\'o}n-Salamanca, A., Merrifield, M., et al.\ 2018, \mnras, 481, 5580
\bibitem[Gao et al.(2018)]{Gao2018} Gao, H., Ho, L.~C., Barth, A.~J., et al.\ 2018, \apj, 862, 100
\bibitem[Gimeno et al.(2004)]{Gimeno2004} Gimeno, G.~N., D{\'{\i}}az, R.~J., \& Carranza, G.~J.\ 2004, \aj, 128, 62
\bibitem[Haynes et al.(2011)]{Haynes2011} Haynes, M.~P., Giovanelli, R., Martin, A.~M., et al.\ 2011, \aj, 142, 170
\bibitem[Ho et al.(2011)]{Ho2011} Ho, L.~C., Li, Z.-Y., Barth, A.~J., et al.\ 2011, \apjs, 197, 21
\bibitem[Huang et al.(2012)]{Huang2012} Huang, S., Haynes, M.~P., Giovanelli    , R., \& Brinchmann, J.\ 2012, \apj, 756, 113
\bibitem[Hubble(1936)]{Hubble1936} Hubble, E. P. 1936, in Realm of the Nebulae, ed. E. P. Hubble (New Haven: Yale Univ. Press), 45
\bibitem[Ishibashi et al.(2013)]{Ishibashi2013} Ishibashi, W., Fabian, A.~C., \& Canning, R.~E.~A.\ 2013, \mnras, 431, 2350
\bibitem[Jogee et al.(2004)]{Jogee2004} Jogee, S., et al.\ 2004, \apj, 615, 105
\bibitem[Kauffmann et al.(2003)]{Kauffmann2003} Kauffmann, G., Heckman, T.~M., Tremonti, C., et al.\ 2003, \mnras, 346, 1055 
\bibitem[Kennicutt(1998)]{K98} Kennicutt, R. C. 1998, \araa, 36, 189
\bibitem[Kennicutt, \& Evans(2012)]{Kennicutt2012} Kennicutt, R.~C., \& Evans, N.~J.\ 2012, \araa, 50, 531
\bibitem[Kewley et al.(2001)]{Kewley2001} Kewley, L.~J., Dopita, M.~A., Sutherland, R.~S., Heisler, C.~A., \& Trevena, J.\ 2001, \apj, 556, 121 
\bibitem[Kormendy(1979)]{Kormendy1979} Kormendy, J.\ 1979, \apj, 227, 714
\bibitem[Kormendy, \& Bender(2012)]{Kormendy2012} Kormendy, J., \& Bender, R.\ 2012, \apjs, 198, 2
\bibitem[Kormendy, \& Kennicutt(2004)]{Kormendy2004} Kormendy, J., \& Kennicutt, R.~C.\ 2004, \araa, 42, 603
\bibitem[Koss et al.(2011)]{Koss2011} Koss, M., Mushotzky, R., Treister, E., et al.\ 2011, \apjl, 735, L42
\bibitem[Kruijssen et al.(2012)]{Kruijssen2012} Kruijssen, J.~M.~D., Pelupessy, F.~I., Lamers, H.~J.~G.~L.~M., et al.\ 2012, \mnras, 421, 1927
\bibitem[Larsen, \& Richtler(1999)]{Larsen1999} Larsen, S.~S., \& Richtler, T.\ 1999, \aap, 345, 59
\bibitem[Lauer et al.(1996)]{Lauer1996} Lauer, T.~R., Tremaine, S., Ajhar, E.~A., et al.\ 1996, \apjl, 471, L79
\bibitem[Lauer et al.(2012)]{Lauer2012} Lauer, T.~R., Bender, R., Kormendy, J., et al.\ 2012, \apj, 745, 121
\bibitem[Laurikainen et al.(2013)]{Laurikainen2013} Laurikainen, E., Salo, H., Athanassoula, E., et al.\ 2013, \mnras, 430, 3489
\bibitem[Laurikainen et al.(2007)]{Laurikainen2007} Laurikainen, E., Salo, H., Buta, R., \& Knapen, J.~H.\ 2007, \mnras, 381, 401 
\bibitem[Lansbury et al.(2014)]{Lansbury2014} Lansbury, G.~B., Lucey, J.~R., \& Smith, R.~J.\ 2014, \mnras, 439, 1749
\bibitem[Lauer et al.(1996)]{Lauer1996} Lauer, T.~R., Tremaine, S., Ajhar, E.~A., et al.\ 1996, \apjl, 471, L79 
\bibitem[Martin et al.(2005)]{Martin2005} Martin, D.~C., Fanson, J., Schiminovich, D., et al.\ 2005, \apjl, 619, L1 
\bibitem[Manzano-King et al.(2019)]{Manzano2019} Manzano-King, C., Canalizo, G., \& Sales, L.~V.\ 2019, arXiv e-prints, arXiv:1905.09287
\bibitem[Mazzarella, \& Boroson(1993)]{Mazzarella1993} Mazzarella, J.~M., \& Boroson, T.~A.\ 1993, \apjs, 85, 27
\bibitem[Mazzarella et al.(2012)]{Mazzarella2012} Mazzarella, J.~M., Iwasawa, K., Vavilkin, T., et al.\ 2012, \aj, 144, 125
\bibitem[Melvin et al.(2014)]{Melvin2014} Melvin, T., Masters, K., Lintott, C., et al.\ 2014, \mnras, 438, 2882
\bibitem[Menezes, \& Steiner(2018)]{Menezes2018} Menezes, R.~B., \& Steiner, J.~E.\ 2018, \apj, 868, 67
\bibitem[Mezcua et al.(2014)]{Mezcua2014} Mezcua, M., Lobanov, A.~P., Mediavilla, E., \& Karouzos, M.\ 2014, \apj, 784, 16 
\bibitem[Mishra et al.(2017)]{Mishra2017} Mishra, P.~K., Barway, S., \& Wadadekar, Y.\ 2017, \mnras, 472, L89
\bibitem[Moiseev et al.(2010)]{Moiseev2010} Moiseev, A., Karachentsev, I., \& Kaisin, S.\ 2010, \mnras, 403, 1849 
\bibitem[Moore et al.(1996)]{Moore1996} Moore, B., Katz, N., Lake, G., et al.\ 1996, \nat, 379, 613
\bibitem[Nordgren et al.(1995)]{Nordgren1995} Nordgren, T.~E., Helou, G., Chengalur, J.~N., et al.\ 1995, \apjs, 99, 461
\bibitem[Nyland et al.(2016)]{Nyland2016} Nyland, K., Young, L.~M., Wrobel, J.~M., et al.\ 2016, \mnras, 458, 2221
\bibitem[Osterbrock(1989)]{Osterbrock1989} Osterbrock, D.~E.\ 1989, Astrophysics of Gaseous Nebulae and Active Galactic Nuclei
\bibitem[Peng et al.(2002)]{Peng2002} Peng, C.~Y., Ho, L.~C., Impey, C.~D., \& Rix, H.-W.\ 2002, \aj, 124, 266 
\bibitem[Peng et al.(2010)]{Peng2010} Peng, C.~Y., Ho, L.~C., Impey, C.~D., \& Rix, H.-W.\ 2010, \aj, 139, 2097 
\bibitem[Penny et al.(2018)]{Penny2018} Penny, S.~J., Masters, K.~L., Smethurst, R., et al.\ 2018, \mnras, 476, 979
\bibitem[Peterson(1979)]{Peterson1979} Peterson, S.~D.\ 1979, \apjs, 40, 527
\bibitem[Poggianti et al.(2017)]{Poggianti2017} Poggianti, B.~M., Moretti, A., Gullieuszik, M., et al.\ 2017, \apj, 844, 48
\bibitem[Portegies Zwart et al.(2010)]{Portegies2010} Portegies Zwart, S.~F., McMillan, S.~L.~W., \& Gieles, M.\ 2010, \araa, 48, 431 
\bibitem[Prieto et al.(2013)]{Prieto2013} Prieto, M., Eliche-Moral, M.~C., Balcells, M., et al.\ 2013, \mnras, 428, 999
\bibitem[Querejeta et al.(2015)]{Querejeta2015} Querejeta, M., Eliche-Moral, M.~C., Tapia, T., et al.\ 2015, \aap, 579, L2
\bibitem[Rizzo et al.(2018)]{Rizzo2018} Rizzo, F., Fraternali, F., \& Iorio, G.\ 2018, \mnras, 476, 2137
\bibitem[Sheth et al.(2008)]{Sheth2008} Sheth, K., Elmegreen, D.~M., Elmegreen, B.~G., et al.\ 2008, \apj, 675, 1141
\bibitem[Sil{\textquoteright}chenko et al.(2019)]{Sil2019} Sil{\textquoteright}chenko, O.~K., Moiseev, A.~V., \& Egorov, O.~V.\ 2019, \apjs, 244, 6
\bibitem[Temi et al.(2009)]{Temi2009} Temi, P., Brighenti, F., \& Mathews, W.~G.\ 2009, \apj, 695, 1
\bibitem[Tremaine(1995)]{Tremaine1995} Tremaine, S.\ 1995, \aj, 110, 628
\bibitem[van den Bergh(1976)]{van1976} van den Bergh, S.\ 1976, \apj, 206, 883
\bibitem[Welch, \& Sage(2003)]{Welch2003} Welch, G.~A., \& Sage, L.~J.\ 2003, \apj, 584, 260
\bibitem[Wilman et al.(2009)]{Wilman2009} Wilman, D.~J., Oemler, A., Mulchaey, J.~S., et al.\ 2009, \apj, 692, 298
\bibitem[Mishra et al.(2019)]{Mishra2019} Mishra, P.~K., Wadadekar, Y., \& Barway, S.\ 2019, arXiv e-prints, arXiv:1905.10739
\bibitem[Xiao et al.(2016)]{Xiao2016} Xiao, M.-Y., Gu, Q.-S., Chen, Y.-M., et al.\ 2016, \apj, 831, 63
\bibitem[York et al.(2000)]{York2000} York, D. G., Adelman, J., Anderson, J. E., Jr., et al. 2000, \aj, 120, 1579   
\bibitem[Yuan et al.(2010)]{Yuan2010} Yuan, T.-T., Kewley, L.~J., \& Sanders, D.~B.\ 2010, \apj, 709, 884
\bibitem[Zhou et al.(2015)]{Zhou2015} Zhou, Z.-M., Cao, C., \& Wu, H.\ 2015, \aj, 149, 1
\bibitem[Zhou et al.(2018)]{Zhou2018} Zhou, Z.-M., Wu, H., Zhou, X., \& Ma, J.\ 2018, \pasp, 130, 094101 
\end{thebibliography}

\end{CJK*}
\end{document}